%% file: main.tex
\documentclass{article}
\usepackage{spconf,amsmath,graphicx, hyperref,xcolor,acronym, comment}

\input{header}

\title{Performance optimizations on deep noise suppression models}
\name{
Jerry Chee$^{\star}$ \qquad 
Sebastian Braun$^{\dagger}$ \qquad
Vishak Gopal$^{\dagger}$ \quad
Ross Cutler$^{\dagger}$}
\address{
$^{\star}$ Cornell University, NY, USA \\
$^{\dagger}$ Microsoft Corporation, Redmond, WA, USA
}

\begin{document}

\maketitle

\begin{abstract}
We study the role of magnitude structured pruning as an architecture search to speed up the inference time of a deep noise suppression (DNS) model.
While deep learning approaches have been remarkably successful in enhancing audio quality, their increased complexity inhibits their deployment in real-time applications.
We achieve up to a 7.25X inference speedup over the baseline, with a smooth model performance degradation.
Ablation studies indicate that our proposed network re-parameterization (i.e., size per layer) is the major driver of the speedup, and that magnitude structured pruning does comparably to directly training a model in the smaller size.
We report inference speed because a parameter reduction does not necessitate speedup, and we measure model quality using an accurate non-intrusive objective speech quality metric.

%Using these insights into the complexity-performance curve, we then present a joint training procedure which enables a single model to adjust its complexity during inference time,
%switching between different model sizes during inference time, 
%without having to store multiple models.
%These nested models can then be used in conjunction with a resource manager to provide flexible levels of noise suppression based on the available computation resources.
%We show a trade-off when deploying a set of models: the memory required to store the set (nested or separately) vs. the resulting quality at each model size.

%Importantly, we do not have to store multiple models.
%A single model is downloaded, and 
%The different models are accessed by taking subsets of the weight tensors. 
%Importantly for real-time applications, no retraining is required when switching between models.
\end{abstract}


\section{Introduction}
There has been much work in compressing deep learning methods so that they can efficiently operate within the real-time and hardware constraints of many audio enhancement applications~\cite{braun21cruse,pqk2021,tan2021compressing,tan2021towards}.
This interest stems from the fact that deep learning methods -- while typically providing superior audio enhancement -- come with a greater computational complexity than classical signal processing methods~\cite{braun21cruse}.
In real-time applications, the computational complexity becomes the primary constraint.
The available memory per device varies, but the available time per computation does not.
Thus we measure and present our compression results in terms of the inference speed.
%Note that 
Calculating the memory or parameter reduction is not an accurate proxy - see Section~\ref{app:sparse}.

We investigate the application of structured pruning and fine-tuning to speed up our baseline CRUSE model~\cite{braun21cruse}.
Structured pruning aims to find a dense sub-network that well approximates the original. 
This type of model compression immediately transfers to an inference speedup 
and reduced storage costs
because we are performing dense and smaller matrix multiplications.
%outputting a dense, smaller, network.
%In contrast, sparse pruning converts dense tensors to sparse ones, and requires specialized software and hardware support to realize any inference speedups or reduced model storage.
%Implementing such sparse support across a meaningfully large section of consumer devices is highly nontrivial.
In addition, we propose a new scalable per-layer parameter configuration for the CRUSE architecture class to specify the pruned network sizes.

%% CASCADE SECTION
%In order to speedup a model we typically compress it to a fixed size.
%%The speedup experiments compress a model to a fixed size.
%%Typically models are compressed to a fixed size. 
%This fixed size can become an issue when many concurrent programs are vying for on-device resources.
%Fixed size models do not ``degrade gracefully''; if they become computationally untenable 
%%on a device 
%they can only be shut off.
%We propose a $N$-cascaded meta-architecture where a flexible complexity level $i \in \{1, \dots, N\}$ can be chosen based on the available computational resources.
%We provide results with the NSnet2~\cite{braun21cruse} model, but the cascade meta-architecture is generic.

\subsection{Contributions}
Using the CRUSE~\cite{braun21cruse} architecture we demonstrate up to a 7.25X speedup over the baseline model, with a smooth degradation of model quality.
%We demonstrate that the CRUSE class of architectures is viable at scales smaller than previously 
%The CRUSE~\cite{braun21cruse} class of noise suppression models are compressed to new levels, and it is shown that their degradation is smooth and the models are still viable.
Ablation studies indicate that the proposed network %re-parameterization 
parameter configuration is in fact responsible for the successful scalability.
Our structured pruning method does no better than directly training a model in a given size. 
The value then of structured pruning is in the architecture search: discovering which network parameterizations can reduce model complexity with minimal model degradation.

%\todo{contributions cascade}

%Real-time background noise suppression
%%audio enhancement
%on consumer devices is both a time
%% computational complexity 
%and memory bound application.
%Deep learning methods have greatly improved noise suppression performance, but at a greater computational complexity cost than classical methods~\cite{braun21cruse}.
%
%faces both computational complexity and memory restrictions.

%%\outline{
%%\begin{itemize}
%%    \item Research trend of real-time, on-device audio enhancement
%%    \item Field is currently lacking in its methodology
%%    \item we bring these improved methodologies
%%    \item we get inference speedup (we can say this bc we actually measure it)
%%    \item Through structured pruning approach, but ablation studies indiciate ...
%%    \item Furthermore, we provide additional scalability for multiple complexity levels
%%\end{itemize}
%%}

\section{Related work}
% compare to ICASSP paper, which had good parameter reduction but no runtime results and only PESQ results (not well correlated to MOS)

Tan and Wang \cite{tan2021compressing,tan2021towards} use sparse regularization, iterative pruning, and clustering-based quantization to compress DNN speech enhancement models. 
However they use STOI and PESQ \cite{p862} to evaluate the quality after compression, which has been shown to have low correlation to subjective quality \cite{cutler2021interspeech,reddy2021interspeech}.
In addition, no run-time benchmarks are given to show real-world improvements, the noise suppression model used is relatively simple and not state of the art, and the training and test set are simplistic. 
Therefore it isn't clear from this study what optimizations work well on a best-in-class noise suppressor on a much more challenging test set such as \cite{reddy2021interspeech}. 

Kim et. al.~\cite{pqk2021} use a combination of unstructured pruning, quantization, and knowledge distillation to compress keyword spotting models.
The authors motivate their work via edge computing, but do not provide any complexity measurements to indicate real-world improvements.
In addition, no comparisons are made to any other compression methods.

% DNS challenge 
% DNS challenge winners
% CRUSE
% 

\subsection{Deep noise suppression}
Braun et. al.~\cite{braun21cruse} developed the CRUSE class of models for real-time deep noise suppression. 
It is based on the U-Net architecture~\cite{tan18unet}, another real-time model for DNS.
%The CRUSE models consist of a $L$ symmetric convolutional and deconvolution encoder and decoder layers, with a parallel GRU layer in between.
%There are also skip connections between each corresponding encoder and decoder layer.
Unlike earlier network architectures primarily based on recurrent neural networks whose models have hit a performance saturation~\cite{weninger15lstm,williamson17tf,xia20weight},
CRUSE is in the class of convolutional recurrent networks~\cite{tan18unet,strake19sep,wichern17low,wisdom19diff}.
These models have increased performance, albeit at a computational cost that prohibits their real-time deployment on consumer devices. We investigate two versions of the CRUSE model, a more complex model called CRUSE32 and a less complex model called CRUSE16.

\subsection{Model compression}
Pruning aims to remove parts of a neural network while maintaining its accuracy.
It can either remove individual parameters resulting in sparse matrices (unstructured)~\cite{frankle2018lottery}, or it can remove groups of parameters such as channels or neurons (structured)~\cite{liu2018rethinking}.
There are many strategies to prune a neural network, but simple magnitude pruning has been shown to be superior to more complicated methods on ImageNet~\cite{hooker19sparse}.
Other methods of model compression include quantization, matrix factorization, and knowledge distillation~\cite{2020compressionsurvey}.
Frankle and Carbin~\cite{frankle2018lottery} present the Lottery Ticket Hypothesis: dense randomly initialized neural networks contain sparse subnetworks (winning tickets) that can be trained to comparable accuracy of the original network, in a comparable number of epochs.
The work of Liu et. al.~\cite{liu2018rethinking} instead focuses on structured pruning, and presents a different message: fine-tuning a pruned model is comparable or worse to training a model directly in that size.

\begin{table}[]
\centering
\begin{tabular}{c|c}
Operator Type & Inference Time Proportion \\
\hline
GRU & 75\% \\
(De)Convolution & 18\% \\
Activations, etc & 7\%
\end{tabular}
\caption{Profiling 
%inference time in 
CRUSE32 to identify main targets for 
\\ inference speedup.}
\label{tab:cruse_profile}
\end{table}

\section{Experimental methodology}
We employ two simple but important experimental methodologies which enable a more accurate assessment of performance in real-world scenarios.
First, we provide timing results in the ONNX Runtime inference engine~\cite{onnxruntime}.
Timing results are crucial because of the strict real-time requirements for background noise suppression.
It is not necessary that reduced parameter counts give a faster inference time.
For example, we measured no meaningful speedup for sparse pruned models in ONNX Runtime (see Section~\ref{app:sparse}).
Benchmark results are conducted on an Intel Core i7-10610U CPU.
Second, the model quality is evaluated using the INTERSPEECH 2021 and ICASSP 2021 DNS Challenge test set~\cite{reddy2021interspeech,reddy2021icassp}.
We use a new non-intrusive objective speech quality metric called DNSMOS P.835 \cite{reddy2021DNSMOSP935} employing the ITU-T P.835~\cite{naderi21p835} standard which provides separate scores for speech (SIG), background (BAK), and overall (OVLR) quality.
%Importantly, DNSMOS P.835 and DNSMOS has been shown to be more reliable in correlating with actual subjective human audio evaluation than PESQ, SDR, and POLQA~\cite{reddy21dnsmos}.
%\todo{talk about how we need better methodology to actually determine if these techniques will be viable in practice, or real-world settings.
%And 2 sub-things: better speech quality evaluament, and more relevant/direct time/complexity measurements.}
DNSMOS P.835 has a Pearson Correlation Coefficient of 0.94 for speech, 0.98 for background, and 0.98 for overall compared to subjective quality, which gives sufficient accuracy to do fast pruning.

%% NOTE: this belongs in \section}{Results}, but placed higher to effect placement in pdf.
\begin{figure*}[]
\centering
\includegraphics[width=0.6\textwidth]{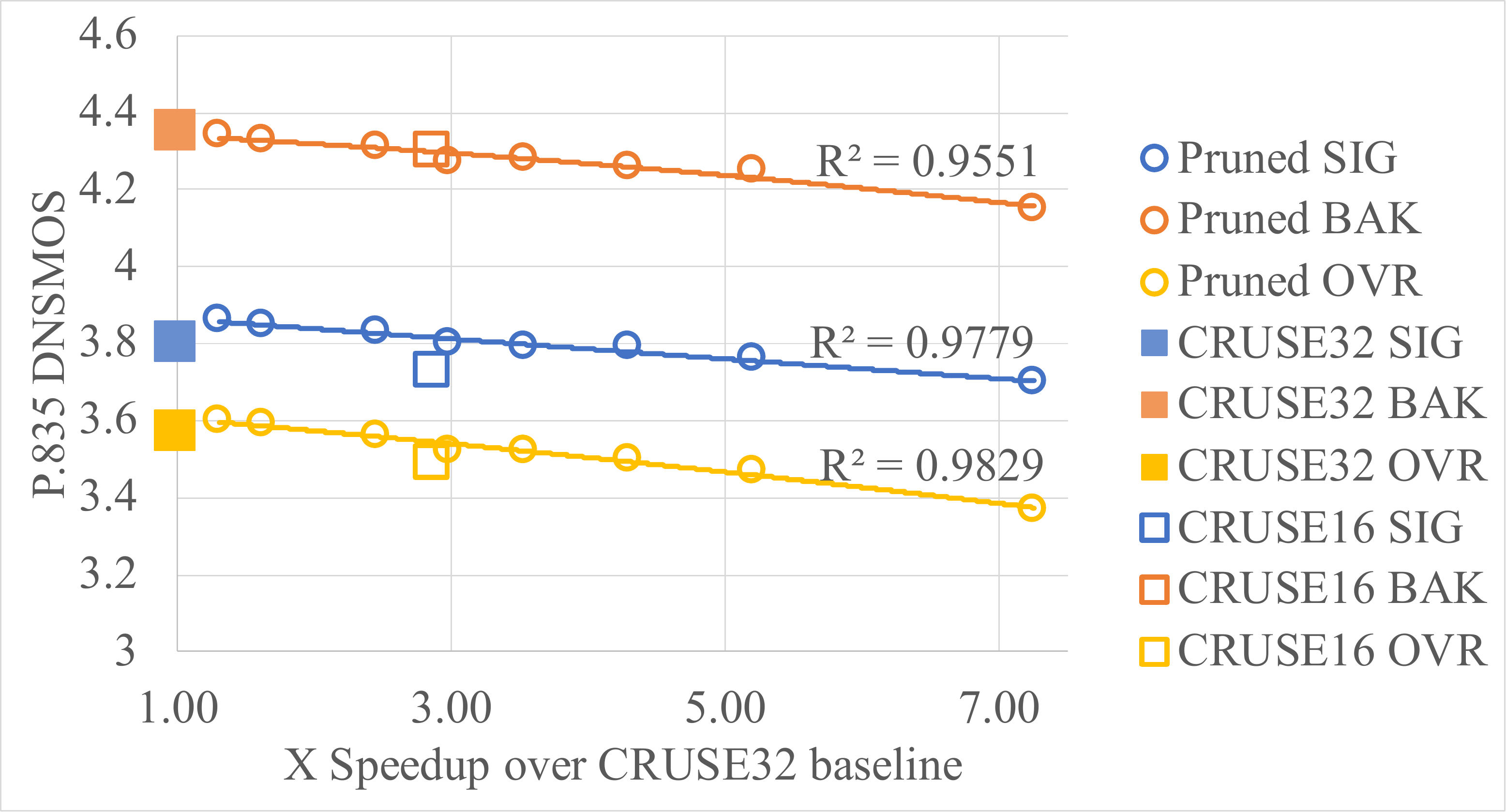}
\caption{Speedup via structured pruning for the CRUSE class of models. 
DNSMOS P.835 reported for the 2021 DNS challenge test sets~\cite{reddy2021interspeech,reddy2021icassp}.
Degree 2 polynomial trend line with $R^2$ values.
}
\label{fig:struct_speedup}
\end{figure*}

%\newpage

\section{Performance optimizations \\via pruning architecture search}
%\section{Structured Pruning as Architecture Search}
%\begin{figure}
%\centering
%\includegraphics[width=0.5\textwidth]{figures/CRUSE.png}
%\caption{\cite{braun21cruse}\todo{placeholder} CRUSE architecture.}
%\label{fig:cruse}
%\end{figure}

We aim to improve the inference time of the CRUSE architecture class while maintaining minimal model degradation.
The CRUSE architecture we study consists of 4 convolution-encoder and deconvolution-decoder layer pairs, with a central parallel GRU layer.
Full details of the architecture can be found in Braun et. al.~\cite{braun21cruse}.
Because the DNS models operate in a real-time environment, we will focus on reducing the inference time.
We first profile the inference time components of CRUSE, with results in Table~\ref{tab:cruse_profile}.
The GRU operations dominate the computation, and so we will focus on compressing them.

\begin{table}[!t]
\centering
\begin{tabular}{l|cccc|c|c}
\multicolumn{1}{c|}{Config} & \multicolumn{4}{c|}{Network} & Mem & Benchmark \\
\multicolumn{1}{c|}{Name} & \multicolumn{4}{c|}{Param} & (MB) & (ms)\\
\hline
%BaselineLrg 
CRUSE32 & 32 & 64 & 128 & 256 & 33.6 & 3.41\hspace{1pt}($\pm$\hspace{1pt}.02) \\ %33.583, 3.407
%BaselineSm 
CRUSE16 & 16 & 32 & 64 & 128 & 8.41 & 1.19\hspace{1pt}($\pm$\hspace{1pt}.02) \\ %8.407 1.192
P.125 & 32 & 64 & 128 & 224 & 26.0 & 2.61\hspace{1pt}($\pm$\hspace{1pt}.01) \\ %26.028 2.609
P.250 & 32 & 64 & 128 & 192 & 19.5 & 2.10\hspace{1pt}($\pm$\hspace{1pt}.01) \\ %19.453 2.095
P.500 & 32 & 64 & 128 & 128 & 9.22 & 1.39\hspace{1pt}($\pm$\hspace{1pt}.02) \\ %9.222 1.385
P.5625 & 32 & 64 & 112 & 112 & 7.16 & 1.14\hspace{1pt}($\pm$\hspace{1pt}.01) \\ % 7.155 1.141
P.625 & 32 & 64 & 96 & 96 & 5.34 & 0.96\hspace{1pt}($\pm$\hspace{1pt}.04) \\ % 5.339 0.963
P.6875 & 32 & 64 & 80 & 80 & 3.79 & 0.79\hspace{1pt}($\pm$\hspace{1pt}.01) \\ %3.794  0.794
P.750 & 32 & 64 & 64 & 64 & 2.52 & 0.65\hspace{1pt}($\pm$\hspace{1pt}.003) \\ % 2.520 0.654
P.875 & 32 & 32 & 32 & 32 & 0.68 & 0.47\hspace{1pt}($\pm$\hspace{1pt}.01) \\ % 0.676 0.470
\end{tabular}
\caption{Structured pruning configurations for CRUSE architecture. 
Model memory is of the ONNX format.
Benchmarks include 95\% confidence intervals.
CRUSE\{32,16\} are baseline models from Braun et. al.~\cite{braun21cruse}.
}
\label{tab:prune_config}
\end{table}

Structured pruning requires that the number of parameters in each layer be adjusted.
Our DNS model can be parameterized by a length 4 vector $[c_1, c_2, c_3, c_4]$.
Because of the skip connections between encoder-decoder layer pairs, CRUSE is symmetric about the central GRU layer.
For example, $c_1$ specifies the output channels of the first convolution layer and the input channels of the second convolution layer.
And $c_4$ specifies the output channels of the last convolution layer and the hidden state of the GRU layer.
Table~\ref{tab:prune_config} specifies the baseline CRUSE32 network parameterizations, as well as the 8 configurations we consider.

We develop a heuristic framework for resizing the CRUSE architecture.
First, reduce the dimensions of the GRU layer via a pruning configuration parameter.
This reduction corresponds to modifying $c_4$ in the network parameterization.
For example, with the P.625 configuration the pruning parameter is $0.625$. 
Thus, we change $c_4 \gets (1-0.625) * 256 = 96$.
Note that we must also change the output channels in the last convolution layer.
Then, we enforce what we call ``network monotonicity'', that $c_1 \leq c_2 \leq c_3 \leq c_4$.
This monotonicity condition is commonly seen in neural network architecture design where the number of channels increase through the network~\cite{vggnet,resnet}.
Recall that $c_3 = 128$, and thus $c_3 > c_4$.  
We then set $c_3 \gets c_4$ to satisfy the condition.
Our proposed configuration scheme can be applied to other U-Net~\cite{tan18unet} style architectures, as it only requires a symmetric encoder-decoder structure.
%network parameterization 
%resizing
%framework 
%utilizes structures common to U-Net architectures~\cite{tan18unet}, and thus is general to other U-Net style architectures with symmetric encoder-decoder structure.

%\begin{table}[]
%\centering
%\begin{tabular}{c|c|c|c}
%Config & Network & Mem & Benchmark \\
%Name & Param & (MB) & (ms)\\
%\hline
%%BaselineLrg 
%CRUSE32 & [32, 64, 128, 256] & 33.583 & 3.407 ($\pm$ .02) \\
%%BaselineSm 
%CRUSE16 & [16, 32, 64, 128] & 8.407 & 1.192 ($\pm$ .02) \\
%P.125 & [32, 64, 128, 224] & 26.028 & 2.609 ($\pm$ .01) \\
%P.250 & [32, 64, 128, 192] & 19.453 & 2.095 ($\pm$ .01) \\
%P.500 & [32, 64, 128, 128] & 9.222 & 1.385 ($\pm$ .02) \\
%P.5625 & [32, 64, 112, 112] & 7.155 & 1.141 ($\pm$ .01) \\
%P.625 & [32, 64, 96, 96] & 5.339 & 0.963 ($\pm$ .04) \\
%P.6875 & [32, 64, 80, 80] & 3.794 & 0.794 ($\pm$ .01) \\
%P.750 & [32, 64, 64, 64] & 2.520 & 0.654 ($\pm$ .003) \\
%P.875 & [32, 32, 32, 32] & 0.676 & 0.470 ($\pm$ .01) \\
%\end{tabular}
%\caption{Structured pruning configurations for CRUSE architecture. 
%CRUSE32 and CRUSE16 are baseline models from Braun et. al.~\cite{braun21cruse}.
%}
%\label{tab:prune_config}
%\end{table}

We employ structured magnitude pruning to construct a dense sub-network to the size specified by our network parameterization.
Consider a generic tensor $W \in \mathbb{R}^{d_1 \times \dots \times d_p}$.
To construct a sub-tensor $\widehat{W}
\in \mathbb{R}^{d_1 \times \dots \times d_{i-1} \times \hat{d}_i \times d_{i+1} \times \dots \times d_p}$ 
with reduced dimension $\hat{d}_i < d_i$, choose a set of coordinates in dimension $i$ satisfying:
\begin{equation*}
%    \hat{W} = \dots
\widehat{\mathcal{J}} = 
\argmax_{\mathcal{J} \subseteq \{1,\dots,d_i\}^{|\hat{d}_i|}} 
\sum_{j \in \mathcal{J}} \| W_i^j \|
\end{equation*}
%
%with $\widehat{W} = $
where $W_i^j$ represents the tensor $W$ indexed at coordinate $j$ in dimension $i$.
We apply this generic structured magnitude pruning method to reduce the number of input or output channels in (de)convolution layers, as well as the input and hidden dimensions of the GRU layer.
We prune the CRUSE32 model, our baseline. 
%as our original.
Fine-tuning is critical to recover model accuracy after pruning.
We use the same optimization hyper-parameters used to train the baseline CRUSE32 model~\cite{braun21cruse}.

\section{Results}

Figure~\ref{fig:struct_speedup} plots the speedup achieved via structured magnitude pruning 
%according 
to the 
%structured pruning
configurations specified in Table~\ref{tab:prune_config}, against the resulting signal, background, and overall DNSMOS P.835.
Importantly, structured pruning 
achieves a smooth trade-off between complexity and model quality.
This trade-off is quite flat; we can decrease complexity without incurring much model degradation.
For example, a 3.64X memory reduction and a 2.46X inference speedup over the baseline CRUSE32 model only incurs a 0.01 degradation in overall DNSMOS P.835.
The P.125 configuration achieves higher DNSMOS P.835 than the baseline CRUSE32 because we are doing additional retraining on top of the baseline model.
At the extreme, a 7.25X inference speedup incurs a 0.2 overall DNSMOS P.835 degradation.
Overall, we have shown the CRUSE class of models is viable at complexity levels previously thought untenable.

\subsection{Ablation studies}
\begin{table}[]
\centering
\begin{tabular}{c|ccc|ccc}
& \multicolumn{6}{c}{DNSMOS P.835} \\
Config & \multicolumn{6}{c}{(SIG, BAK, OVRL)} \\
Name & \multicolumn{3}{c|}{Struct Prune} & \multicolumn{3}{c}{Direct Train} \\
\hline
P.125 & 3.86 & 4.34 & 3.60 
& 3.85 & 4.33 & 3.59 \\
P.250 & 3.85 & 4.33 & 3.59 
& 3.85 & 4.34 & 3.59 \\
P.500 & 3.83 & 4.31 & 3.56 
& 3.82 & 4.32 & 3.56 \\
P.750 & 3.76 & 4.25 & 3.47 
& 3.77 & 4.26 & 3.48 \\
P.875 & 3.70 & 4.15 & 3.37 
& 3.68 & 4.17 & 3.36 
\end{tabular}
\caption{Prune vs direct training a configuration.}
\label{tab:direct_v_prune}
\end{table}

We have introduced two new variables in compressing the CRUSE class of models: the network re-parameterization, and the structured magnitude pruning method itself.
The CRUSE16 model follows the previous network parameterization~\cite{braun21cruse}, but with half the number of parameters as CRUSE32.
In Table~\ref{tab:prune_config} we see that CRUSE16 lies between the P.500 and P.5625 prune configurations in terms of memory and benchmark speed.
However Figure~\ref{fig:struct_speedup}
shows CRUSE16 has a worse signal and overall DNSMOS P.835, with equivalent background.
Holding 
%network
complexity constant, our new network parameterization in Table~\ref{tab:prune_config} achieves superior audio quality. \\
%as described
%This result shows the value of our proposed network parameterization in enabling complexity reduction with minimal model degradation.

Table~\ref{tab:direct_v_prune} shows the results of directly training a model in a given configuration, using the same total number of training epochs as the pruned models.
The results are effectively indistinguishable, which indicates that structured magnitude pruning is not providing any added value.
%\todo{Question: should I run a t-test?}
Additional tuning of the fine-tuning procedure for structured pruning did not improve results, see Table~\ref{tab:tune_lr_struct}.

\begin{table}[t]
\centering
\begin{tabular}{l|ccc}
\multicolumn{1}{c|}{Init} & \multicolumn{3}{c}{DNSMOS P.835 } \\
\multicolumn{1}{c|}{LR} & \multicolumn{3}{c}{(SIG, BAK, OVRL)} \\
\hline
1e-3* & 3.76 & 4.25 & 3.47 \\
1e-4 & 3.70 & 4.18 & 3.39 \\
1e-5 & 3.61 & 4.08 & 3.24 \\
1e-6 & 3.46 & 3.80 & 3.04 \\
\end{tabular}
\caption{Tuning LR for structured pruning fine-tuning.
(*)~is the selected hyper-parameter, used for fine-tuning, direct training, and is the original CRUSE32 training parameter.
}
\label{tab:tune_lr_struct}
\end{table}

%\todo{CI for DNSMOS?}

\begin{table}[t]
\centering
\begin{tabular}{c|ccc}
& \multicolumn{3}{c}{DNSMOS P.835} \\
Baseline & \multicolumn{3}{c}{(SIG, BAK, OVR} \\
\hline
CRUSE32 
& 3.80 & 4.35 & 3.57 \\
CRUSE16
& 3.73 & 4.30 & 3.49
\end{tabular}
\caption{
DNSMOS P.835 for baseline CRUSE32, CRUSE16 models.
}
\label{tab:cruse_baseline}
\end{table}

\subsection{The value of fine-tuning}
%\subsection{Structured pruning before and after fine-tuning}
\begin{table}[!t]
\centering
\begin{tabular}{l|ccc|ccc}
 & \multicolumn{6}{c}{DNSMOS P.835} \\
 & \multicolumn{6}{c}{(SIG, BAK, OVRL)} \\
Prune & \multicolumn{3}{c|}{Before} & \multicolumn{3}{c}{After} \\
Name & \multicolumn{3}{c|}{Fine-Tuning} & \multicolumn{3}{c}{Fine-Tuning} \\
\hline
P.125 & 2.22 & 2.87 & 2.11 & 3.86 & 4.34 & 3.60 \\
P.250 & 2.23 & 2.87 & 2.15 & 3.85 & 4.33 & 3.59 \\
P.500 & 2.03 & 2.83 & 1.96 & 3.83 & 4.31 & 3.56 \\
P.5625 & 2.04 & 2.77 & 2.04 & 3.80 & 4.27 & 3.52 \\
P.625 & 2.11 & 2.74 & 2.02 & 3.79 & 4.28 & 3.52 \\
P.6875 & 2.21 & 2.81 & 2.00 & 3.79 & 4.26 & 3.50 \\
P.750 & 2.02 & 2.75 & 1.89 & 3.76 & 4.25 & 3.47 \\
P.875 & 2.66 & 2.74 & 2.22 & 3.70 & 4.15 & 3.37 
\end{tabular}
\caption{DNSMOS P.835 of structure pruned CRUSE32 models before and after fine-tuning.}
\label{tab:struct_FT}
\end{table}

\begin{table}[t]
\centering
\begin{tabular}{c|ccc|ccc}
 & \multicolumn{6}{c}{DNSMOS P.835 } \\
 & \multicolumn{6}{c}{(SIG, BAK, OVRL)} \\
Frac & \multicolumn{3}{c|}{Before} & \multicolumn{3}{c}{After}\\
GRU & \multicolumn{3}{c|}{Fine-Tuning} & \multicolumn{3}{c}{Fine-Tuning} \\
\hline
%0.00 
%& 3.80 & 4.35 & 3.57 \\
0.25
& 3.80 & 4.34 & 3.56 
& 3.86 & 4.32 & 3.59 \\
0.50
& 3.78 & 4.32 & 3.54 
& 3.86 & 4.33 & 3.60 \\
0.75 
& 3.68 & 4.22 & 3.40 
& 3.86 & 4.33 & 3.59 \\
\end{tabular}
\caption{
DNSMOS P.835 of unstructured (sparse) pruned CRUSE32 models before and after fine-tuning.
}
\label{tab:sparse_FT}
\end{table}

In Table~\ref{tab:struct_FT} the DNSMOS P.835 is reported after structured magnitude pruning, and again after fine-tuning.
We see that fine-tuning is crucial to recover the model quality. 
The MOS scores are increased by approximately 1, a large margin.
Table~\ref{tab:sparse_FT} reports the DNSMOS P.835 after unstructured (sparse) magnitude pruning, and again after fine-tuning.
Remarkably, we see that unstructured pruning does not degrade the model quality nearly as much.
Note, the ``Frac GRU 0.25'' unstructured setting 
does not modify any convolution layers and 
is not equivalent to the ``P.250'' configuration.
%because in the unstructured setting we are not modifying any of the convolution layers.
Subsequently, fine-tuning has less to recover.

\subsection{Sparsity does not imply inference speedup}
%\subsection{Sparse pruning results}
\label{app:sparse}
\begin{table}[t]
\centering
\begin{tabular}{c|c}
Frac & Benchmark \\
GRU & (ms) \\
\hline
0.00 & 4.91 ($\pm$ .07) \\ %4.907
0.25 & 4.92 ($\pm$ .04) \\ %4.919
0.50 & 4.92 ($\pm$ .08) \\ %4.923
0.75 & 4.90 ($\pm$ .03) \\ %4.896
\end{tabular}
\caption{
Benchmark results of unstructured (sparse) magnitude pruning the GRU layers in CRUSE32.
}
\label{tab:sparse}
\end{table}

Despite the promising results in Table~\ref{tab:sparse_FT}, reducing the parameter count via sparsity does not easily translate to an inference speedup.
Table~\ref{tab:sparse} shows the benchmark results for sparse magnitude pruning of the GRU layer in CRUSE32. 
A certain fraction of the GRU weights were set to zero.
The times are not distinguishable with their confidence intervals.
This means that ONNX Runtime provides no %significant 
speedup from sparse linear algebra operations.
Specialized sparse inference support is needed to realize speedup from sparsity.
The Neural Magic inference engine~\cite{neuralmagic} is one such option, but in our experimentation we found that support for RNN layers was still in development.

\section{Conclusions}
We have achieved up to a 7.25X inference speedup over the baseline CRUSE32 model.
Our experiments indicate that the proposed network parameterization (size of each layer) is the primary driver of our speedup results, not structured magnitude pruning.
These conclusions support Liu et. al.~\cite{liu2018rethinking} in that the value of structured pruning is in conducting an architecture search. 
Additionally, our methodological choice of measuring the inference speed -- rather than using parameter count as a proxy -- revealed the difficulty of realizing practical benefits from sparse pruning methods.
Our proposed network parameterization only requires a symmetric encoder-decoder structure, and thus can be applied to other U-Net style architectures.

\newpage
\small
\bibliographystyle{IEEEbib}
\bibliography{refs}
%\printbibliography

%\newpage
%\appendix
\end{document}

%% file: header.tex
\usepackage{amsmath}
\usepackage{amssymb}
\usepackage{amsthm}
\usepackage[linesnumbered,ruled,vlined,noend]{algorithm2e}
\usepackage{subcaption}
\usepackage{multirow}
\usepackage{booktabs}
\usepackage{xcolor}

\usepackage{spconf,graphicx}
\usepackage{hyperref}
\hypersetup{
    colorlinks=true,
    linkcolor=blue,
    filecolor=magenta,      
    urlcolor=cyan,
}
\DeclareMathOperator*{\argmax}{argmax}